\documentclass[twocolumn, twocolappendix]{aastex7}

\usepackage{amsmath}
\usepackage{amssymb}
\usepackage{bm}
\usepackage{xcolor}

\received{October 5, 2026}

\submitjournal{AJ}

\newcommand{\parens}[1]{\left(#1\right)}
\newcommand{\brackets}[1]{\left[#1\right]}

\graphicspath{{./}{figs/}}
\begin{document}

\title{Optimal Photon Counting with Fast, Low Noise Astronomical Imagers\footnote{This paper includes data gathered with the 6.5 meter Magellan Telescopes located at Las Campanas Observatory, Chile. }}

\author[0000-0002-6401-778X]{Jack T. Dinsmore}
\email{jtd@stanford.edu}
\affiliation{Department of Physics, Stanford University, Stanford, CA 94305, USA}
\affiliation{Kavli Institute for Particle Astrophysics and Cosmology, Stanford University, Stanford, MA 94305, USA}

\author[0000-0002-7191-4403]{Christopher Layden}
\email{clayden7@mit.edu}
\affiliation{Department of Physics, Massachusetts Institute of Technology, Cambridge, MA 02139, USA}
\affiliation{Kavli Institute for Astrophysics and Space Research, Massachusetts Institute of Technology, Cambridge, MA 02139, USA}

\author[0000-0001-6711-3286]{Roger W. Romani}
\email{rwr@stanford.edu}
\affiliation{Department of Physics, Stanford University, Stanford, CA 94305, USA}
\affiliation{Kavli Institute for Particle Astrophysics and Cosmology, Stanford University, Stanford, MA 94305, USA}

\author[0000-0002-7303-8144]{Vera L. Berger}
\email{vlberger@mit.edu}
\affiliation{Department of Physics, Massachusetts Institute of Technology, Cambridge, MA 02139, USA}
\affiliation{Kavli Institute for Astrophysics and Space Research, Massachusetts Institute of Technology, Cambridge, MA 02139, USA}

\author[0000-0002-7226-836X]{Kevin B. Burdge}
\email{kburge@mit.edu}
\affiliation{Department of Physics, Massachusetts Institute of Technology, Cambridge, MA 02139, USA}
\affiliation{Kavli Institute for Astrophysics and Space Research, Massachusetts Institute of Technology, Cambridge, MA 02139, USA}

\author[0000-0001-6331-112X]{Geoffrey Mo}
\email{gmo@caltech.edu}
\affiliation{The Observatories of the Carnegie Institution for Science, 813 Santa Barbara St., Pasadena, CA 91101, USA}
\affiliation{Department of Astronomy, California Institute of Technology, 1216 E California Blvd, Pasadena, CA 91125, USA}

\author[0000-0001-8804-8946]{Deepto Chakrabarty}
\email{deepto@mit.edu}
\affiliation{Department of Physics, Massachusetts Institute of Technology, Cambridge, MA 02139, USA}
\affiliation{Kavli Institute for Astrophysics and Space Research, Massachusetts Institute of Technology, Cambridge, MA 02139, USA}

\author[0000-0003-4780-4105]{Emma T. Chickles}
\email{echickle@mit.edu}
\affiliation{Department of Physics, Massachusetts Institute of Technology, Cambridge, MA 02139, USA}
\affiliation{Kavli Institute for Astrophysics and Space Research, Massachusetts Institute of Technology, Cambridge, MA 02139, USA}

\author[0000-0001-8467-9767]{Gabor Furesz}
\email{gfuresz@mit.edu}
\affiliation{Kavli Institute for Astrophysics and Space Research, Massachusetts Institute of Technology, Cambridge, MA 02139, USA}

\author[0000-0003-1361-985X]{Juliana García-Mejía}
\email{jgarciam@mit.edu}
\affiliation{Department of Physics, Massachusetts Institute of Technology, Cambridge, MA 02139, USA}
\affiliation{Kavli Institute for Astrophysics and Space Research, Massachusetts Institute of Technology, Cambridge, MA 02139, USA}
\affiliation{Center for Astrophysics \textbar\ Harvard \& Smithsonian, 60 Garden Street, Cambridge, MA 02138, USA}

\author[0000-0003-0412-9664]{Dave Osip}
\email{dosip@carnegiescience.edu}
\affiliation{Las Campanas Observatory, Carnegie Institution for Science, Colina el Pino, Casilla 601 La Serena, Chile}

\author[0000-0002-9602-2217]{Jack Piotrowski}
\email{piotrowski.jack@gmail.com}
\affiliation{The Observatories of the Carnegie Institution for Science, 813 Santa Barbara St., Pasadena, CA 91101, USA}

\begin{abstract}
    Measuring rapid astrophysical phenomena requires many short exposures, which suffer from repeated read noise. A variety of recent technologies are achieving sub-electron read noise, which enable rapid time-domain studies of faint sources. These data can be analyzed with traditional methods, but the ability to resolve individual electrons enables a new class of more precise, photon-counting techniques. We present a photon-counting method for sub-electron read noise detectors which delivers optimal statistical uncertainties.  We focus on Complementary Metal-Oxide-Semiconductor (CMOS) detectors and demonstrate our method on optical data collected by CMOS instrument ``proto-Lightspeed,'' recently commissioned at the Magellan Clay telescope at Las Campanas Observatory. The method can be applied to non-CMOS detectors as well, such as skipper CCDs and linear-mode avalanche photodiodes. The method substantially improves analysis of faint sources, increasing signal to noise ratios by up to $\sim 130\%$, which would require 5.3$\times$ more observing time to accomplish with data alone. As an example we show detections of four normal rotation-powered pulsars with confirmed optical pulses, achieved in under two hours of total exposure. We provide the \texttt{lightspeedpy} pipeline for reducing proto-Lightspeed data with both photon counting and standard techniques.
\end{abstract}

\keywords{instrumentation: detectors, methods: data analysis, pulsars: individual (B0531+21, B0540$-$69, B0833$-$45, B0656+14)}

\section{Introduction}
Rapid cadence data can provide valuable information regarding many astrophysical objects. The ingress and egress of exoplanet transits and stellar occultations by solar system bodies constrain the objects' physical shape, atmosphere, and other properties. Pulsars, X-ray binaries, and other compact objects can have sub-second period light curves and non-periodic flares, which constrain the objects' energy budgets and emission mechanisms. The rapid fluctuations of Earth's atmosphere can also be diminished or removed with a rapid imager using lucky imaging or speckle interferometry.

However, exposures short enough to measure this rapid variability are typically photon starved. Low read noise is thus required to accurately measure rapid variability for faint objects. Pulsars provide a particularly severe case: their short periods require $\sim$ms exposures to resolve the pulse, but with optical magnitude of $V \gtrsim 22$  only a few photons enter the telescope within each interval (except the $V=16.6$ Crab pulsar). Consequently only four normal rotation-powered pulsars have confirmed optical magnetospheric pulsations: Crab (PSR B0531+21), its LMC twin (PSR B0540$-$69, hereafter B0540), Vela (PSR B0833$-$45), and PSR B0656+14, hearafter B0656 \citep{cocke1969discovery, middleditch1985optical, wallace1977detection, shearer1997pulsed}. Unconfirmed pulsations have also been detected for Geminga \citep{shearer1998possible}. Two millisecond pulsars (MSP) also show optical modulation, which may represent surface emission or interaction with the striped pulsar wind: the transitional MSP J1023+0038 and the MSP SAX J1808.4$-$3658 \citep{ambrosino2017optical, ambrosino2021optical}. Such studies often use specialized detectors such as avalanche photodiode detectors (e.g., Iqueye \cite{naletto2009iqueye} and Optima \cite{straubmeier2001optima}), or electron multiplying charge-coupled devices (EMCCDs), which can achieve sub-electron read noise.

\begin{figure*}
    \centering\leavevmode
\raisebox{-\height}{\includegraphics[width=0.49\linewidth]{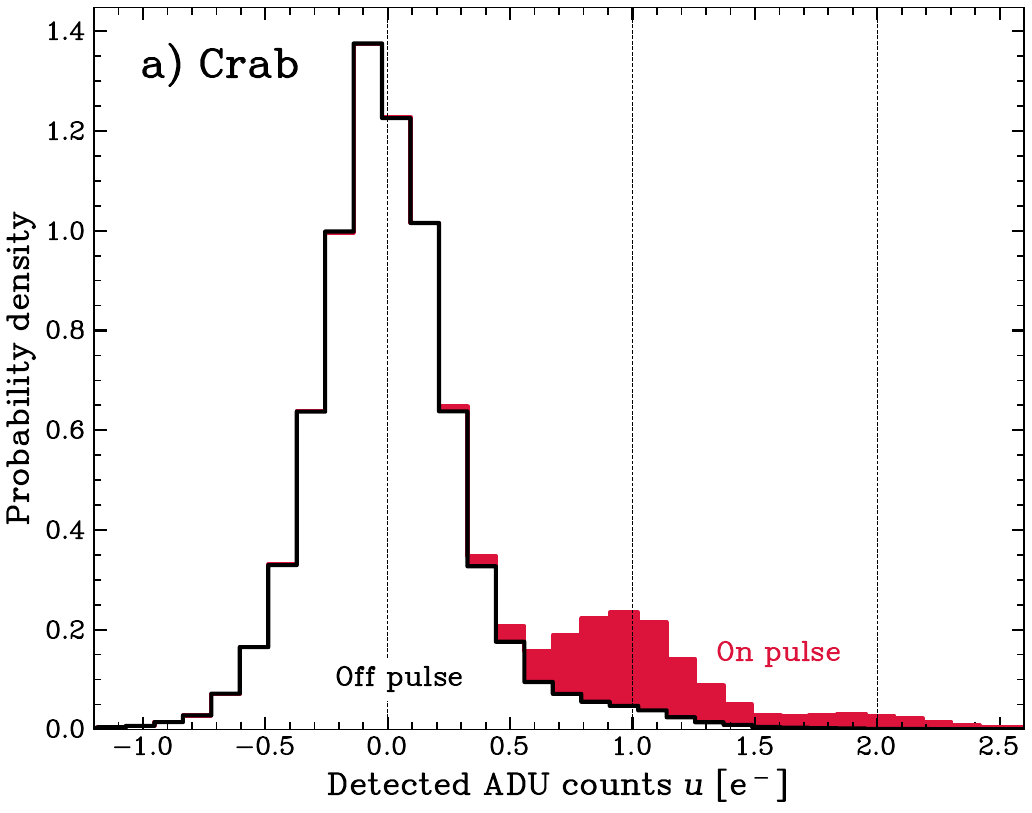}}%
\raisebox{-\height}{\includegraphics[width=0.49\linewidth]{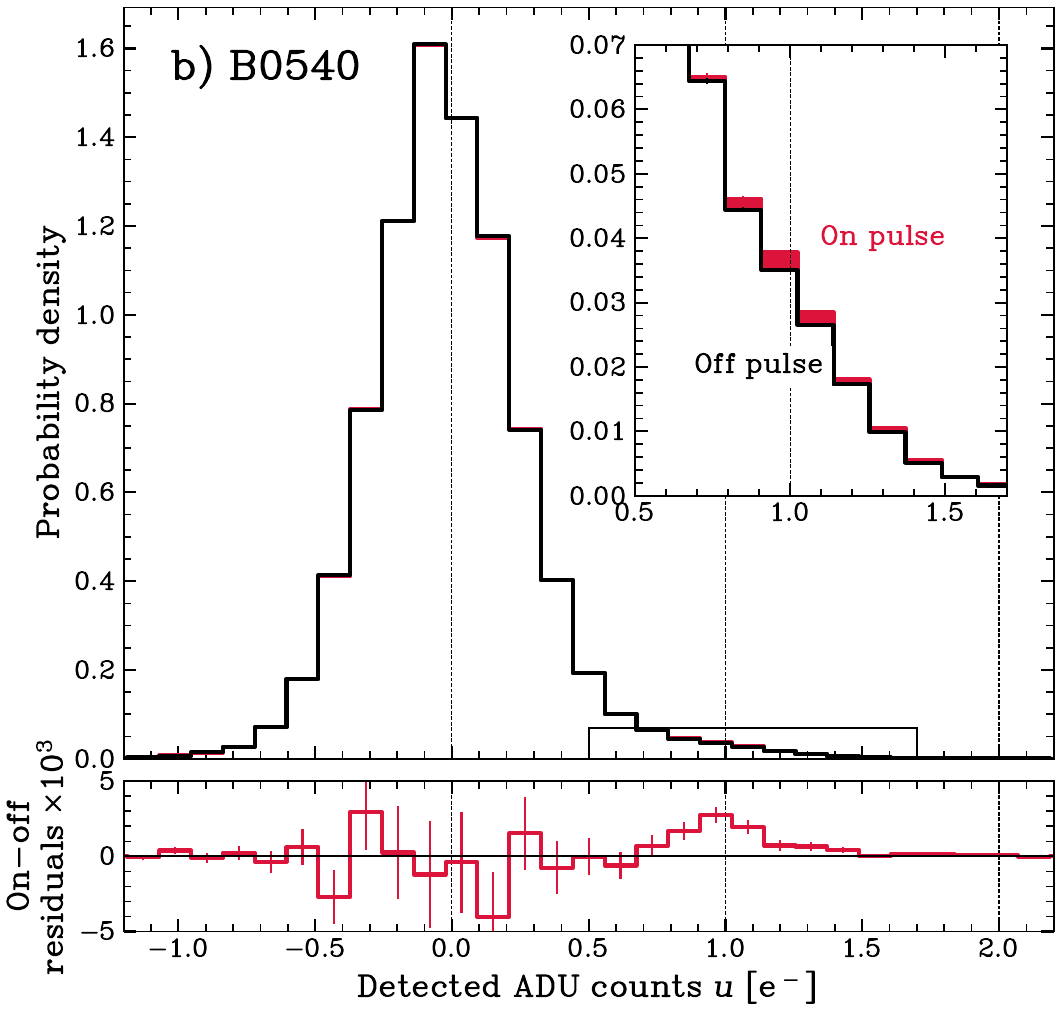}}
    \caption{%
        Distribution of ADU counts (converted to electrons) in the ``on'' and ``off'' phases of \textbf{a}) the Crab pulsar and  \textbf{b}) PSR B0540$-$69, observed with proto-Lightspeed. The inset in b) shows a zoom-in on the boxed portion of the distribution's tail. In the case of the Crab, the photoelectron peaks can clearly be seen during the pulse and are resolved by the ORCA-Quest 2 camera's $\sim0.3$ e$^-$ read noise. All techniques can detect this signal. In the case of the much fainter PSR B0540$-$69, the 1 e$^-$ peak is subdominant compared to the noise tail. The ``weight'' technique is needed to best extract this signal.
        \label{fig:distro}
    }
\end{figure*}

Recent developments have enabled complementary metal-oxide-semiconductor (CMOS) detectors to rival other detectors with rapid readout and low read noise, making them an excellent candidate for studying rapid time-domain astronomy. For example, the proto-Lightspeed instrument was recently commissioned on the Magellan Clay telescope at Las Campanas Observatory \citep{layden2026proto}. Its CMOS detector achieves sub-electron read noise ($\sim 0.3$ e$^-$) and can image at windowed frame rates $>6$ kHz, although it demonstrates reduced quantum efficiency (QE) and nonlinear response at the lowest fluxes \citep{layden2025characterization,layden2026proto}.

For imagers such as proto-Lightspeed, faint photometry is often performed by summing the camera's raw output count rate within a source aperture, without attempting to remove read noise. But when read noise is $\sigma \lesssim 1$ e$^-$, peaks appear in the distribution of detected photoelectrons (see e.g.~the red ``on pulse'' histogram of Fig.~\ref{fig:distro}a) which enable more advanced techniques. A first approach is to round pixel values to integer numbers of detected electrons. This ``rounding'' method can be superior to the former ``summing'' method, especially when the rounding thresholds are adjusted \citep{tulloch2011on, harpsoe2012bayesian, teranishi2012required, fossum2013modeling}. Nevertheless it struggles when the source flux is small enough that the noise fluctuations above 1 e$^-$ dominate the photoelectron signal, as these fluctuations could be interpreted as signal electrons (see Fig.~\ref{fig:distro}b). However, one can fit a model to the histogram of pixel values. By fitting, one effectively increases the weight of pixel reads which resemble a detected photon while de-weighting reads that resemble noise fluctuations. This ``weight'' method delivers more precise best-fit result because it is controlled by the most constraining pixel reads, whereas the ``sum'' or ``round'' results treat all pixel reads equally even when the read value is inconsistent with a photoelectron detection. However, this ``weight'' method requires an accurate model for the detector's noise and flux-dependent QE to avoid biased flux estimators.

In this work, we give a recipe for reducing data with the ``weight'' method. \S\ref{sec:methods} describes the technique, proving that it delivers optimal uncertainties in most cases and is superior to the ``sum'' and ``round'' methods at all values of read noise. These results are applicable to any non-electron-multiplying, sub-electron read noise detector, such as skipper CCDs or linear-mode avalanche photodiodes. For EMCCDs, the technique could be modified to incorporate multiplication noise \citep[see e.g.][]{harpsoe2012bayesian}.
\S\ref{sec:lightspeed} extracts the calibration products necessary to use the model on proto-Lightspeed data, appropriately accounting for the detector's flux-dependent QE. A pipeline for reducing proto-Lightspeed data using the ``sum,'' ``round,'' and ``weight'' methods is also introduced, named \texttt{lightspeedpy}. Despite the method's apparent complexity, ``weight'' analyses require only slightly more processing time than standard analyses with \texttt{lightspeedpy}. \S\ref{sec:results} applies the ``weight'' method to the four rotation-powered pulsars with detected optical pulsations, demonstrating improved significance compared to the other methods. We conclude with \S\ref{sec:conclusion}. 

\section{The ``Weight'' Method}\label{sec:methods}
The raw output from each pixel is the number of detected analog-to-digital units (ADU). We divide by the detector conversion gain (in ADU/e$^-$) to obtain the read value $u$ in units of electrons. Each pixel's $u$-value is drawn from a probability mass function (PMF) $P(u | \lambda)$, where $\lambda$ is the average number of photons incident on the pixel per frame.\footnote{Note that $\lambda$ denotes the incident photon rate \textit{after} passing through the instrument and telescope. $\lambda$ should be divided by quantum efficiencies, throughputs, and flat fields after the fit to obtain the source flux.} The weight method determines $\lambda$ by fitting $P(u | \lambda)$ to the histogram of the $u$-values. \S\ref{sec:parametrization} gives a parameterization of $P(u | \lambda)$ so that this fit function's shape can be calibrated using bias frames. \S\ref{sec:fit} describes how the fit is then performed.

\subsection{Parameterization of the Model}\label{sec:parametrization}
Each exposure, an integer number of photons $\gamma$ interact with a given pixel, where $\gamma$ is Poisson-distributed:
\begin{equation}
    P(\gamma | \lambda) = \frac{e^{-\lambda} \lambda^\gamma}{\gamma!}.
\end{equation}
This function sets the initial heights of the histogram peaks shown in Fig.~\ref{fig:distro}a. These photons eject photoelectrons, but detector defects may prevent some of these photons from being detected, which leads to a flux-dependent internal QE. (We refer to this electron detection inefficiency as the ``internal'' QE to distinguish it from the QE reduction incurred by photodiode-fill factor or absorption/reflection by the photon detection region.) Suppose that the number of detected photoelectrons is $\epsilon \leq \gamma$, distributed by the PMF $P(\epsilon | \gamma)$. This function encapsulates the detector's flux-dependent internal QE and alters the heights of the histogram peaks. Finally, the camera reports $u$ electrons detected, shifted from $\epsilon$ due to the detector's read noise. We write the noise distribution as $P(u | \epsilon)$, which smooths the histogram into a continuous distribution. Altogether, the model function $P(u | \lambda)$ satisfies
\begin{equation}
    P(u | \lambda) = \sum^\infty_{\gamma=0} \sum_{\epsilon = 0}^\gamma P(u|\epsilon) P(\epsilon | \gamma) P(\gamma | \lambda).
    \label{eqn:like}
\end{equation}
This distribution may be different for each pixel, especially for CMOSs which have multiple amplifiers and ADCs, each with different noise and internal QE properties.

If several thousand bias frames are collected, the zero-flux ADU count distribution is well measured in each pixel. We refer to this bias frame distribution (divided by the conversion gain) as $B(u)$, which is equal to $P(u | \epsilon=0)$ by definition. For the rapid readout detectors we consider here, collecting thousands of bias frames poses little issue. We approximate the noise distribution at larger $\epsilon$ by shifting the bias distribution up by $\epsilon$; i.e.\,$P(u | \epsilon) = B(u - \epsilon)$. This assumption may imperfectly characterize the tails of the $P(u|\epsilon \geq 1)$ distribution, but these tails are far less important than the measured $P(u | \epsilon=0)$ tails; for faint sources nearly all pixels have $\epsilon = 0$, and for bright sources the details of the noise distribution are washed out by shot noise.

As $P(\epsilon | \gamma)$ should depend on camera architecture, we discuss it in the context of proto-Lightspeed's CMOS camera in \S\ref{sec:lightspeed}. If $B(u)$ and $P(\epsilon | \gamma)$ are known, then the model function $P(u | \lambda)$ can be evaluated. For long exposures, electrons from dark current also contribute to $\epsilon$ \citep[0.005 e$^-$ s$^{-1}$ pixel$^{-1}$ for proto-Lightspeed,][]{layden2026proto}, which Eq.~\ref{eqn:like} must properly take into account. At present, we assume that dark current electrons suffer the same internal QE as photoelectrons. Further assuming that the dark electrons are Poisson distributed, their effect can be modeled as an additive offset to $\lambda$ which is subtracted with a dark frame (as is conventional for ``sum''-method analyses.) We adopt this approach here, but caution that the dark current statistics may need more detailed treatment for long exposure data from non-cryogenic cameras with substantial dark current, such as proto-Lightspeed.

\subsection{Fit Method} \label{sec:fit}
For the sake of simplicity, we describe the new ``weight'' method for extracting light curves. The procedure can be straightforwardly modified to imaging or other applications as discussed at the end of this section.

Once the desired number of light curve bins has been determined, the object is to estimate the source count rate in each bin. We refer to this rate as $\lambda_b$ in photons per pixel per frame, where $b$ indexes over bins. For each bin and each pixel $p$ in the source's region of interest (ROI), we collect all the measured $u$-values into histograms. We refer to the counts in histogram bin $u$ as $n_{u,pb}$. These $n_{u,pb}$ counts are multinomial distributed, where bin $u$ has probability $P_p(u|\lambda_b)$. The subscript $p$ serves to distinguish between the $P(u|\lambda_b)$ distributions of different pixels. The likelihood for the full data set is
\begin{equation}
    L = \prod_p \prod_b \parens{\sum_u n_{u,pb}}!\prod_u \frac{P_p(u|\lambda_b)^{n_{u,pb}}}{n_{u,pb}!}.
\end{equation}
The maximum likelihood estimator (MLE) for $\lambda_b$ can be found by maximizing $L$, or, equivalently, $\ln L$. Removing the terms which do not depend on $\lambda_b$, the function to maximize is $\ln L = \sum_{u,p,b} n_{u,pb} \ln P_p(u|\lambda_b)$.

For numerical efficiency, we perform this maximization using Newton's method. An initial guess for $\lambda_b$ is formulated using the ``round'' method and then iteratively updated by $\lambda_b \rightarrow \lambda_b - \sum_{b'} g_{b'} (H^{-1})_{bb'}$, where $g_{b'}$ and $H_{bb'}$ are the gradient and Hessian matrix of $\ln L$ with respect to $\lambda_b$:
\begin{align}
    g_b &= \sum_{u,p,b} n_{u,pb} \frac{P_p'(u|\lambda_b)}{P_p(u|\lambda_b)}\\
    H_{bb'} &= \delta_{bb'} \sum_{u,p,b} n_{u,pb} \brackets{\frac{P_p''(u|\lambda_b)}{P_p(u|\lambda_b)} - \parens{\frac{P_p'(u|\lambda_b)}{P_p(u|\lambda_b)}}^2},
\end{align}
where primes indicate derivatives with respect to $\lambda_b$, and $\delta_{bb'}$ is the Kronecker delta. Each $\lambda_b$ evaluation is rapid because these derivatives are analytic, and typically $< 20$ iterations are required before the $\lambda_b$ shift is less than one part in 10$^3$, at which point we identify the fit as converged. The analysis speed is in fact comparable to the unweighted ``sum'' and ``round'' methods.

MLE methods allow uncertainties to be estimated by calculating the second derivative of $\ln L$. But instead, we calculate the uncertainty of $\lambda_b$ using bootstrapped errors, which better capture possible unmodeled effects such as unmasked cosmic rays. Bootstrapped errors are calculated by re-sampling the data set many times (with replacement), recomputing the fluxes for each sample, and reporting the standard deviation of the resampled flux values. By default, \texttt{lightspeedpy} employs 16 bootstrap iterations.

This procedure can be straightforwardly generalized to extract other data products. For example, PSF-weighted light curves could be extracted by treating the bin probability ($P_p(u|\lambda_b)$ above) as $P_p(u|I_p\lambda_b)$, where $I_p$ is the PSF intensity at pixel $p$. Note that the $P_p'$ and $P_p''$ derivatives introduce factors of $I_p$ and $I_p^2$, since they should be taken with respect to $\lambda_b$. Background subtraction is further included by introducing a new parameter $\lambda_\mathrm{bkg}$, and using bin probability $P_p(u|I_p\lambda_b + \lambda_\mathrm{bkg})$. Frames may also be stacked to form an image using a weighted method. Here, one collects a histogram of $u$-values $n_{u,p}$ for each pixel, and the fit parameters are the count rate of each pixel, $\lambda_p$.

\subsection{Statistical Optimality of the Method}
In the limit of large data sets, the MLE is the minimum variance unbiased estimator. While the data are initially the set of pixel reads $\{u\}$, our choice to reduce to histograms $\{n_{u,pb}\}$ loses no information in most cases because each read is independent. Our method then reports the MLE resulting from a fit to the histogram, making it the statistically optimal method (though the method's precision and accuracy could be further improved with upgraded noise and flux-dependent QE models).

\begin{figure}
    \centering
    \includegraphics[width=0.8\linewidth]{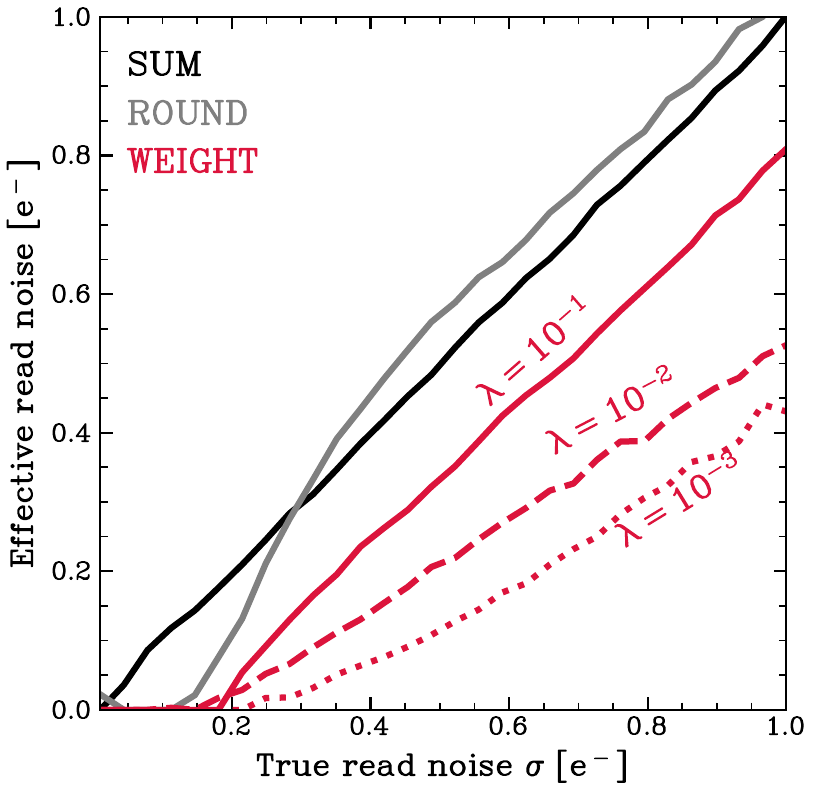}
    \caption{%
        Simulated flux uncertainty delivered by various methods as a function of Gaussian detector read noise. The solid, dashed, and dotted lines denote differing values of source flux ($\lambda$, counts per pixel per frame). The weight method (described in this work) is superior to the ``sum'' method (adding all pixel values), and superior or equal to the ``round'' method (rounding each pixel value to the nearest electron) at all values of true read noise. It also achieves zero effective read noise at larger values of true read noise than the other methods.
        \label{fig:theory}
    }
\end{figure}

Fig.~\ref{fig:theory} compares the effective read noise as a function of true read noise between various methods. The ``sum'' method estimates flux by summing the $u$ data, the ``round'' method first rounds $u$ to the nearest electron, and the ``weight'' method is proposed above. The data displayed in this figure considers a theoretical linear detector with Gaussian noise of standard deviation $\sigma$. The solid lines assume a flux of $\lambda = 10^{-1}$ counts per pixel. For the ``weight'' method, the dashed/dotted lines use $\lambda = 10^{-2} / 10^{-3}$. The flux dependence of the ``sum'' and ``round'' methods is much weaker, so we do not show dashed/dotted lines for those methods. The effective read noise is determined by subtracting $\sqrt{\lambda}$ shot noise in quadrature from the flux uncertainty (measured via bootstrapping).

As expected, the ``sum'' method's effective read noise is simply the detector's true read noise. The ``round'' method is inferior to the standard ``sum'' method for read noise near unity, but becomes effective at $\sigma \sim 0.3$ e$^-$ \cite[also found by][]{teranishi2012required, fossum2013modeling, layden2026proto}. ``Round'' achieves zero effective read noise at $\sigma \sim 0.15$ e$^-$, once the electron peaks become well separated. In contrast to ``round,'' the ``weight'' method is superior to the ``sum'' method at all values of read noise, including in the $\sigma > 0.3$ regime. The ``weight'' method also achieves the zero effective read noise earlier than the ``round'' method. These benefits are exaggerated at lower fluxes, as illustrated by the dotted line. This demonstrates that the ``weight'' method is increasingly efficient for faint sources. One would expect the methods to converge to the same effective read noise at large values of $\sigma$, and we confirmed this fact numerically for $\sigma \gg 1$.

There are some exceptions to the optimality of the histogram approach. For example we assumed that each frame contributes to only one light curve bin $b$. However some frames will contribute to multiple bins, and because we ignore this effect, the choice to reduce the data to $n_{u,pb}$ histograms does lose information. Properly treating the contribution of each frame to multiple bins would allow further detail to be extracted at time scales shorter than the exposure time. This concept generalizes to other applications: for example, consider extracting a stacked image by applying shifts to each frame to correct for atmospheric variation. Reducing the data to $n_{u,p}$ histograms requires shifts to be integer numbers of pixels. Treating each frame-pixel as potentially contributing to multiple stacked-pixels would allow sub-pixel shifts, sharpening the image. These inefficiencies can be addressed if the histogram step is abandoned, and the likelihood is written purely in terms of the $u$-value read from each pixel. While feasible, this approach comes at increased complexity and computational cost, and we do not implement it in this work.

\section{Proto-Lightspeed Detector Modeling}
\label{sec:lightspeed}

\begin{figure}
    \centering
    \includegraphics[width=\linewidth]{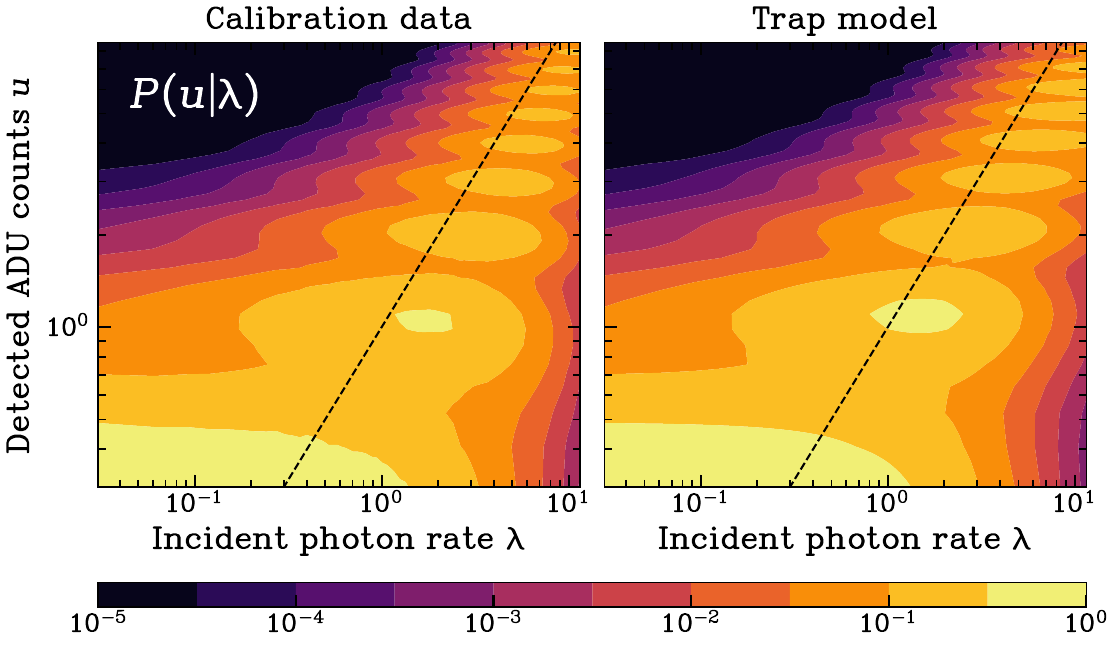}
    \caption{%
        The distribution of detected ADU as a function of the incident photon rate $P(u|\lambda)$, as measured from calibration data with all pixels combined (\textit{left}) and predicted from the model discussed in \S\ref{sec:lightspeed} (\textit{right}).  The diagonal dotted line represents an ideal detector ($u = \lambda$), so the fact that the majority of the distribution lies at smaller $u$-values illustrates the detector's flux-dependent QE. The agreement between the calibration data and the model demonstrates that our detector systematic modeling is accurate enough to employ the weight method.
        \label{fig:likelihood}
    }
\end{figure}

This section discusses the noise distribution $P(u|\epsilon)$ and the internal QE function $P(\epsilon | \gamma)$ in the context of proto-Lightspeed's ORCA-Quest 2 camera, manufactured by Hamamatsu. This camera employs the HWK4123 CMOS image sensor from Fairchild Imaging \citep{cho_2023}.

The accuracy of our detector characterization is demonstrated by Fig.~\ref{fig:likelihood}. The left panel depicts the fit function $P(u|\lambda)$ measured from calibration data and described further in \S\ref{sec:nonlinearity}. The right panel depicts the prediction for $P(u|\lambda)$ generated using the methods described below. The good match between the two demonstrates that the weight method is sufficiently calibrated to reduce data with minimal bias. In particular, the model  is always within a factor of two of the true distribution at $\lambda < 1$ (where the weighting scheme is most useful).

\subsection{Noise Distribution}
We model the proto-Lightspeed noise distribution as a triple Gaussian with identical variances but distinct amplitudes and means (six variables in total). A triple Gaussian noise distribution arises naturally from the output of a source follower transistor if the transistor has one carrier trap near its Si/SiO$_2$ interface and is read out via correlated double sampling \citep[CDS][]{wang:2006}. In this case, signal readings in the non-central Gaussians correspond to instances in which the trap occupation changed between correlated samples. This behavior in image sensors is often referred to as random telegraph noise (RTN). It was previously reported that $> 3.9\%$ of pixels in the proto-Lightspeed image sensor have this triple-Gaussian noise distribution, including a large fraction of the pixels with larger-than-average read noise \cite{Layden_2026_rtn}. Though the majority of proto-Lightspeed pixels likely do not suffer from RTN, triple Gaussians yielded far better fits to the noise distributions than single Gaussians for nearly all pixels. We do not seek a full physical understanding of the noise distribution $B(u)$ in each pixel but rather optimal parametrization of this distribution, so we use a triple Gaussian fit for all pixels. Improved understanding of the sensor noise or application to other devices might prefer different noise models.

After collecting several thousand bias frames, we fit these six parameters to the $B(u)$ bias-frame distribution for each pixel in parallel using gradient descent with an adaptive learning rate \citep{barzilai1988two}. Investigation of $B(u)$ shows that each pixel has a set of ADU values that are never detected \citep[this was first reported by][]{gallagher_2024}. We exclude these values when performing fits for each pixel, to avoid bias. This model provides a good description of pixel outputs when using the camera's ``ultra-quiet'' readout mode, except for a small set of pixels with unusually large read noise. As these ``bad'' pixels are de-weighted by the weight method, their poor noise model is a negligible effect. We do not characterize the alternative, ``standard'' observing mode, which delivers larger read noise. 

\begin{figure}
    \centering
    \includegraphics[width=0.85\linewidth]{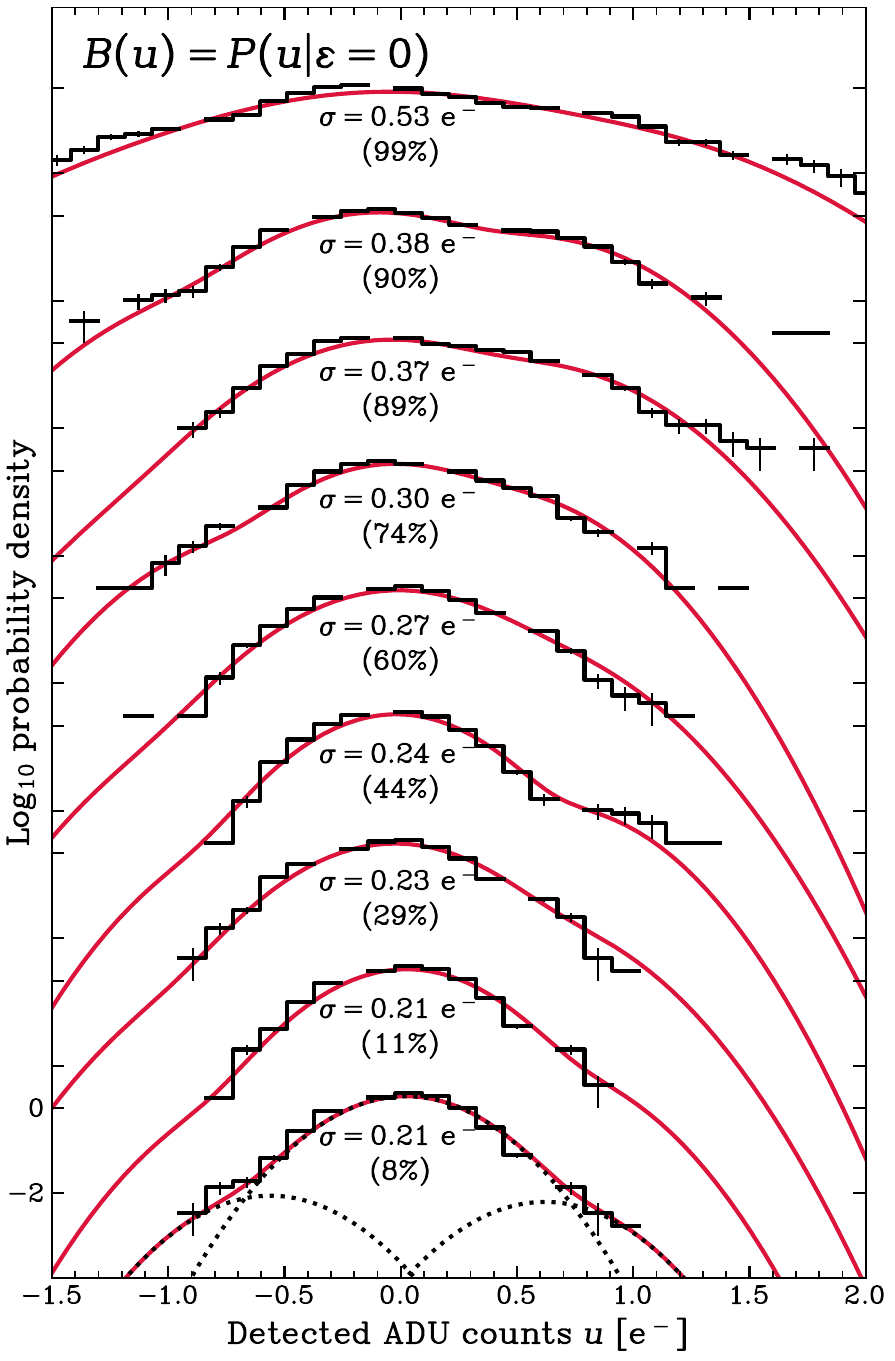}
    \caption{%
        Proto-Lightspeed $B(u)$ noise distributions measured from bias data and shown for randomly selected pixels. The distributions are listed in decreasing RMS noise $\sigma$ from top to bottom, and the percentile of the noise value is given in parentheses. The curves are shifted vertically for visual clarity. The best fit curves are shown as red lines. Dotted lines show the composite Gaussians for the bottommost distribution. Note the strong pixel-to-pixel variation, which requires each pixel to be separately modeled for observations of very faint sources. In the median pixel, approximately 3\% of reads fluctuate by $>0.5$ e$^-$ and would be confused with a true photoelectron if not treated with the weighting method.
        \label{fig:noise}
    }
\end{figure}

For the instrument's ultra-quiet mode, the most common pixel read noise value was $0.22$ e$^-$ with the majority of pixels showing higher values. In the fitted models, the central Gaussian is generally far larger amplitude than the other two model components, which are separated by $\sim 0.5$ e$^-$ \citep[in agreement with][]{Layden_2026_rtn}. Example noise distributions are shown in Fig.~\ref{fig:noise}. The gaps in the black histograms represent the unrealized ADU values excluded from the fits.

\subsection{Internal Quantum Efficiency}
\label{sec:nonlinearity}

\begin{figure}
    \centering
    \includegraphics[width=\linewidth]{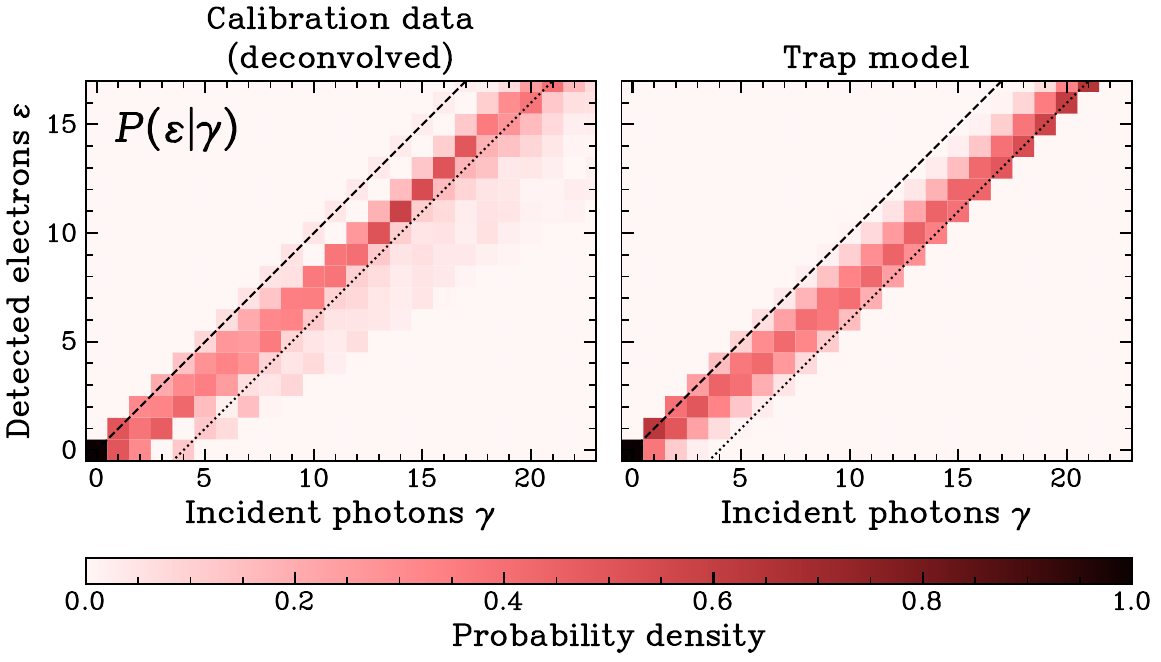}
    \caption{%
        The PMF for the number of detected electrons as a function of the number of incident photons, $P(\epsilon | \gamma)$. \textit{Left}: The function deconvolved from calibration data of all pixels; \textit{right}: the ``electron trap'' model prediction (see text). The two panels show similar features, such as a tendency for $\epsilon = \gamma-4$ at large $\gamma$ (dotted line), which suggests that there are four electron traps on average.
        \label{fig:efficiency}
    }
\end{figure}

The function $P(\epsilon | \gamma)$ alters the photoelectron peaks in the $u$ histogram. This function is closely related to the internal QE, which is the average ratio between $\epsilon$ and $\lambda$ a function of $\lambda$. However, the full distribution of $P(\epsilon | \gamma)$ is necessary for the weighted analysis, not just this ratio.

We first attempt to measure $P(\epsilon | \gamma)$ as a function of $\epsilon$ and $\gamma$ directly from the calibration data shown in Fig.~\ref{fig:likelihood}. These data were collected by exposing a bright target with an exponentially increasing exposure time $t$ to measure the distribution of detected ADU counts $u$ as a function of the incident photon rate $\lambda \propto t$. The constant of proportionality was determined by assuming that proto-Lightspeed's response is linear at large fluxes, as found by \cite{layden2026proto}, and by using a calibrated photodiode. The data from all of proto-Lightspeed's 9.4 megapixels per frame are merged, allowing for one very high signal map of the pixel-averaged $P(u | \lambda)$ distribution.

For this analysis we restrict to the low exposure data sets with $0.5 < \lambda < 23$, since the internal QE is most flux-dependent at low count rate. Accordingly, we restrict the domain to $\epsilon \leq 24$ and $\gamma \leq 30$ only, as larger values are not constrained by the $\lambda < 23$ data. Since $\epsilon$ and $\gamma$ are integers, $P(\epsilon | \gamma)$ can be expressed within the chosen domain using 775 numbers; including the constraints that $\epsilon \leq \gamma$ and $\sum_\epsilon P(\epsilon | \gamma) = 1$ yields 444 total fit parameters. The model function $P(u | \lambda)$ is fitted to the data by minimizing the $\chi^2$ statistic using \texttt{scipy.optimize.least\_squares}. The resulting $P(\epsilon | \gamma)$ distribution is shown in the left panel of Fig.~\ref{fig:efficiency}. The fit results with $\epsilon \geq 18$ and $\gamma \geq 24$ are not shown as they are prone to unphysical edge effects in the numerical solution.

This fit process is similar to a deconvolution procedure. Like deconvolution, it tends to amplify small perturbations to the modeled noise distribution, and we find that the results depend on the $\epsilon, \gamma,$ and $\lambda$ cuts. We are also accumulating calibration data from many pixels and likely missing pixel-to-pixel variations in the $P(\epsilon | \gamma)$ distribution. This fit result is therefore not accurate enough to be used as the final $P(\epsilon | \gamma)$ distribution. Instead, we use the result to motivate an analytical model for CMOS flux-dependent internal QE, which will be the final model for $P(\epsilon | \gamma)$.

\cite{layden2025characterization} suggest an ``electron trap'' model for CMOS flux-dependent internal QE, in which a photoelectron can be trapped in the CMOS depletion layer by a local electric potential minimum, preventing it from being detected. When a large number of photoelectrons are present, all traps are likely to be filled, leaving $\epsilon = \gamma - T$ for $T$ traps. Our data trend towards $\epsilon = \gamma - 4$ at large $\gamma$ (dotted line), consistent with the trap model if $T=4$ for the average pixel. The low$-\gamma$ portion of data is best modeled when each unoccupied trap has probability $p\approx 11\%$ to catch a photoelectron. The trap model prediction for $P(\epsilon | \gamma)$ for $p=11\%$ and $T=4$ is shown in the right hand side of Fig.~\ref{fig:efficiency}. The model follows the general trend of the calibration data, so we adopt these parameters for the final $P(\epsilon | \gamma)$ distribution. However the data appears more smoothed at large $\gamma$. This could occur if the number of traps $T$ varies pixel-to-pixel, which would smear the calibration data distribution along the vertical axis. We expect such variation to occur due to the pixel-to-pixel internal QE variations reported by \cite{layden2026proto}, but we do not treat them in the present work.

\subsection{\texttt{lightspeedpy}: an Analysis Package for proto-Lightspeed}
\label{sec:lightspeedpy}
We introduce the \texttt{lightspeedpy} package to perform sum, round, and weight analysis of proto-Lightspeed data. This includes documented utilities to create calibrated light curves, photometry, and stacked images. Algorithms to process calibration data and recover the $P(u | \lambda)$ distribution are also provided, including methods to extract the noise distribution $P(u|\epsilon)$ and the QE function $P(\epsilon | \gamma)$. The $T=4$ and $p=11\%$ QE parameters are employed by default, though these values can be adjusted. The package can be used as a command line utility, or imported as a Python programming interface for customized analyses. The proto-Lightspeed data discussed in the following section were reduced with \texttt{lightspeedpy}.

The code is publicly available on Github\footnote{\url{https://github.com/jtdinsmore/lightspeedpy.git}} where it will be maintained by the authors. It is designed as a ``plugin system,'' which allows community members to add or modify utilities on their local installations and suggest changes to the main package via Github pull requests, which we encourage users to do. The current release is available on Zenodo \citep{dinsmore2026lightspeedpy}.

\section{Results}
\label{sec:results}
As pulsars are very fast and optically faint, their light curves represent an excellent use case for precise analysis methods for low-noise detectors. This section displays phase-folded light curves of the four rotation-powered pulsars with confirmed optical pulses --- Crab, B0540, Vela, and B0656 --- from proto-Lightspeed test data. Previous studies have analyzed their light curves in detail, and we do not uncover any new features \citep[our B0656 curve however has higher phase resolution than published results by][]{kern2003optical}. Accordingly we do not discuss source physics here and instead use the data to evaluate the fit methods.

\subsection{Observations}
On September 12, 2025, we observed the Crab with proto-Lightspeed using 1.3 ms exposures. 24 seconds of that observation are shown in this work. On December 18, 2025, we observed B0656 with 10 ms exposures with 1.2 hours total exposure. On December 19, 2025, we observed B0540 and Vela with 3 ms exposures, with 6.3 and 25 minutes total exposure respectively. No filters were used, so observations occupy proto-Lightspeed's unfiltered band of $\sim 400-700$ nm. As proto-Lightspeed biases vary on $\sim$hour timescales, we collected biases immediately before each observation (and interleaved biases within the longer B0656 observation) to ensure the noise distribution was well characterized. We read out a strip of the detector to achieve high frame rates. In some cases we were able to orient this strip to simultaneously cover a bright comparison star. This allowed us to stack the frames in post-processing and align the comparison star to remove pointing drift.

All observations were collected with the ultra-quiet mode to obtain the lowest possible read noise.  Though proto-Lightspeed's ORCA-Quest 2 camera provides a photon-counting mode, we do not use it as that mode is analogous to the ``round'' method discussed above, which is not as sensitive as the ``weight'' method we apply here. We employ proto-Lightspeed's maximum pixel scale of 0$''$.05.

The nights yielded good seeing ($\sim 0''.5$), but we encountered image stability issues due to unexpected freedom of movement in one of proto-Lightpseed's commercial reimaging lenses. This error led an internal component of the lens to slip unpredictably, shifting the image. In the case of B0656, we accounted for this shift by comparison star tracking, but for Vela no field star was available, and the integrated image quality is much worse.

\begin{figure}
    \centering
    \includegraphics[width=0.988\linewidth]{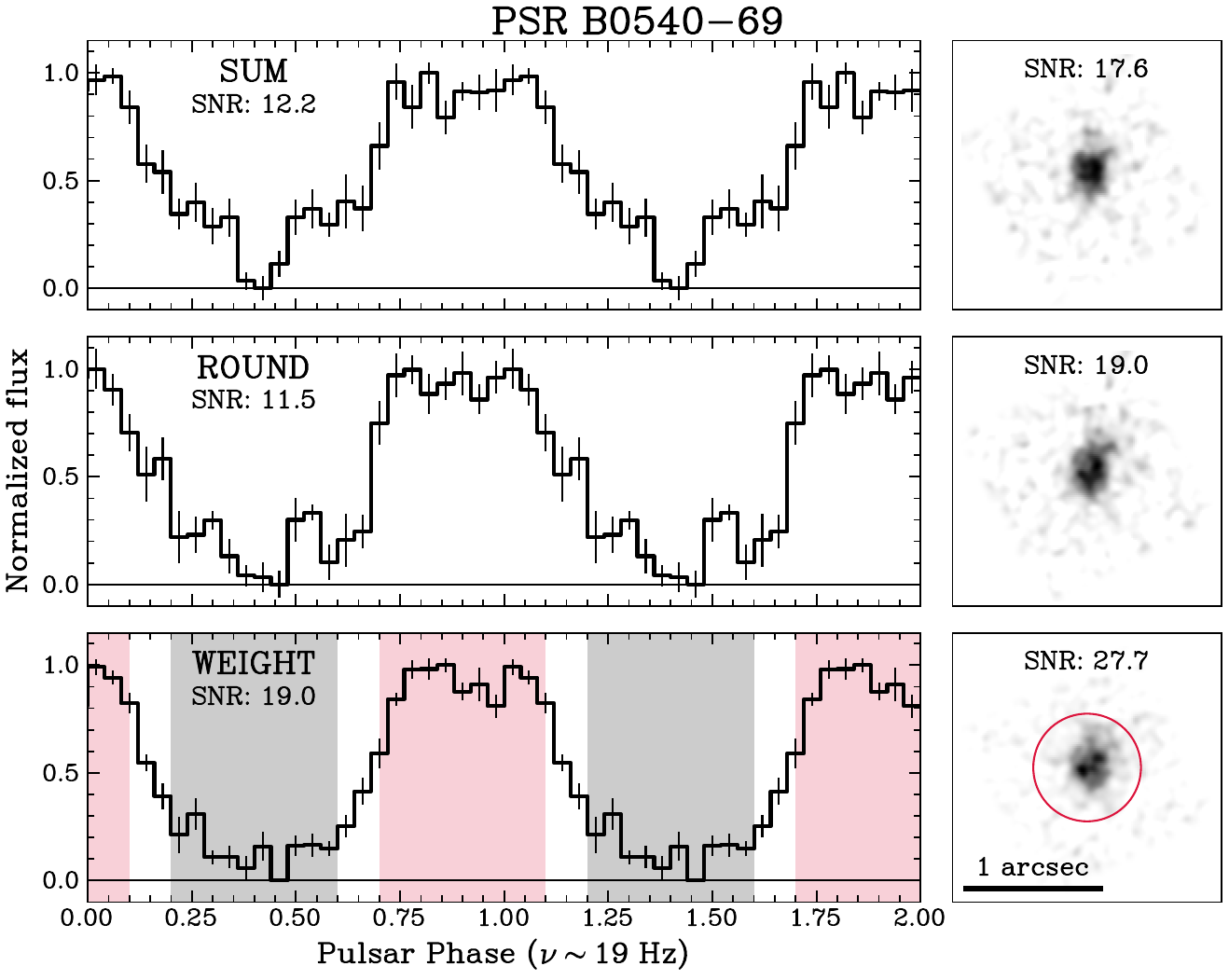} \vspace{0.18cm}

    \includegraphics[width=0.988\linewidth]{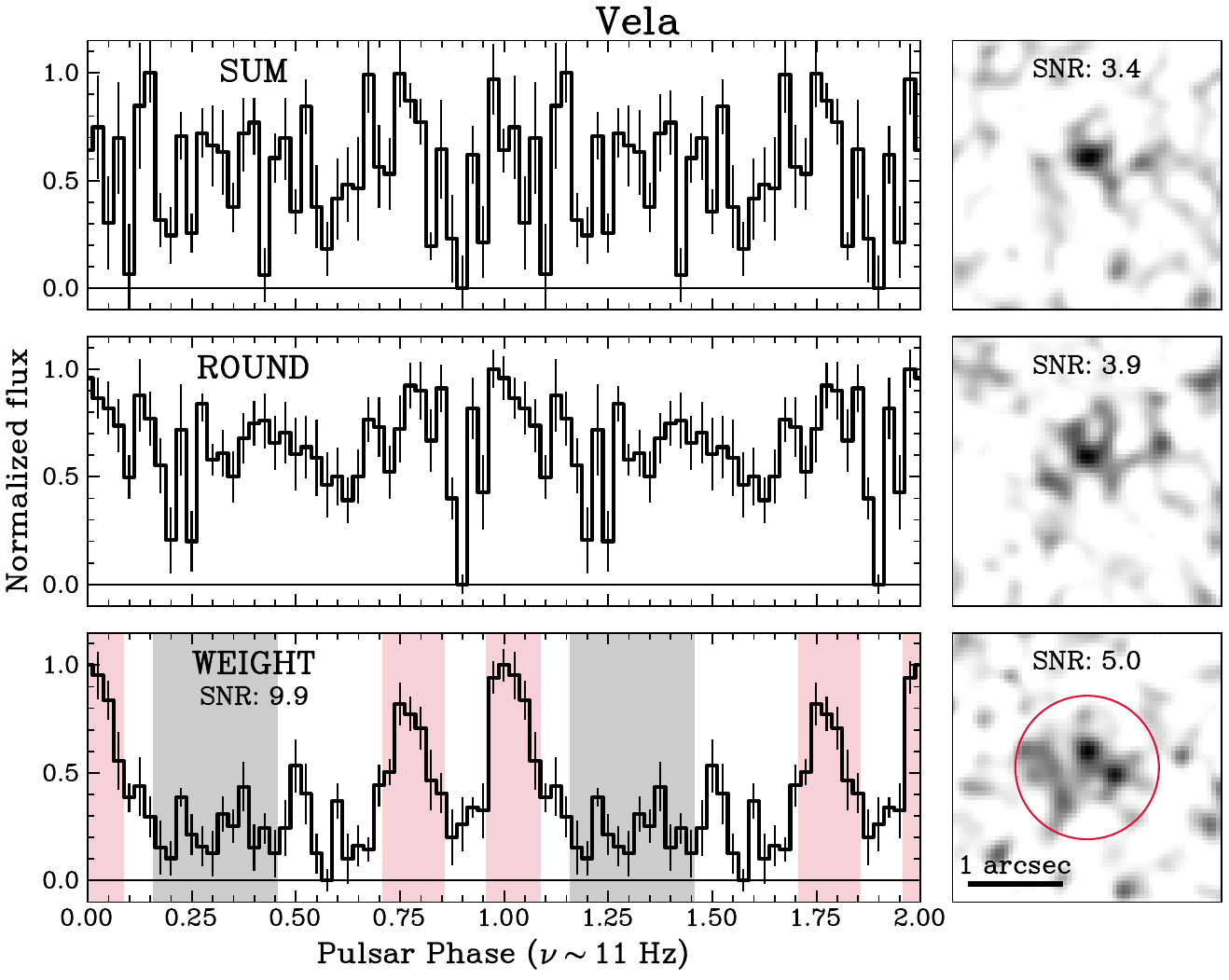} \vspace{0.18cm}

    \includegraphics[width=0.988\linewidth]{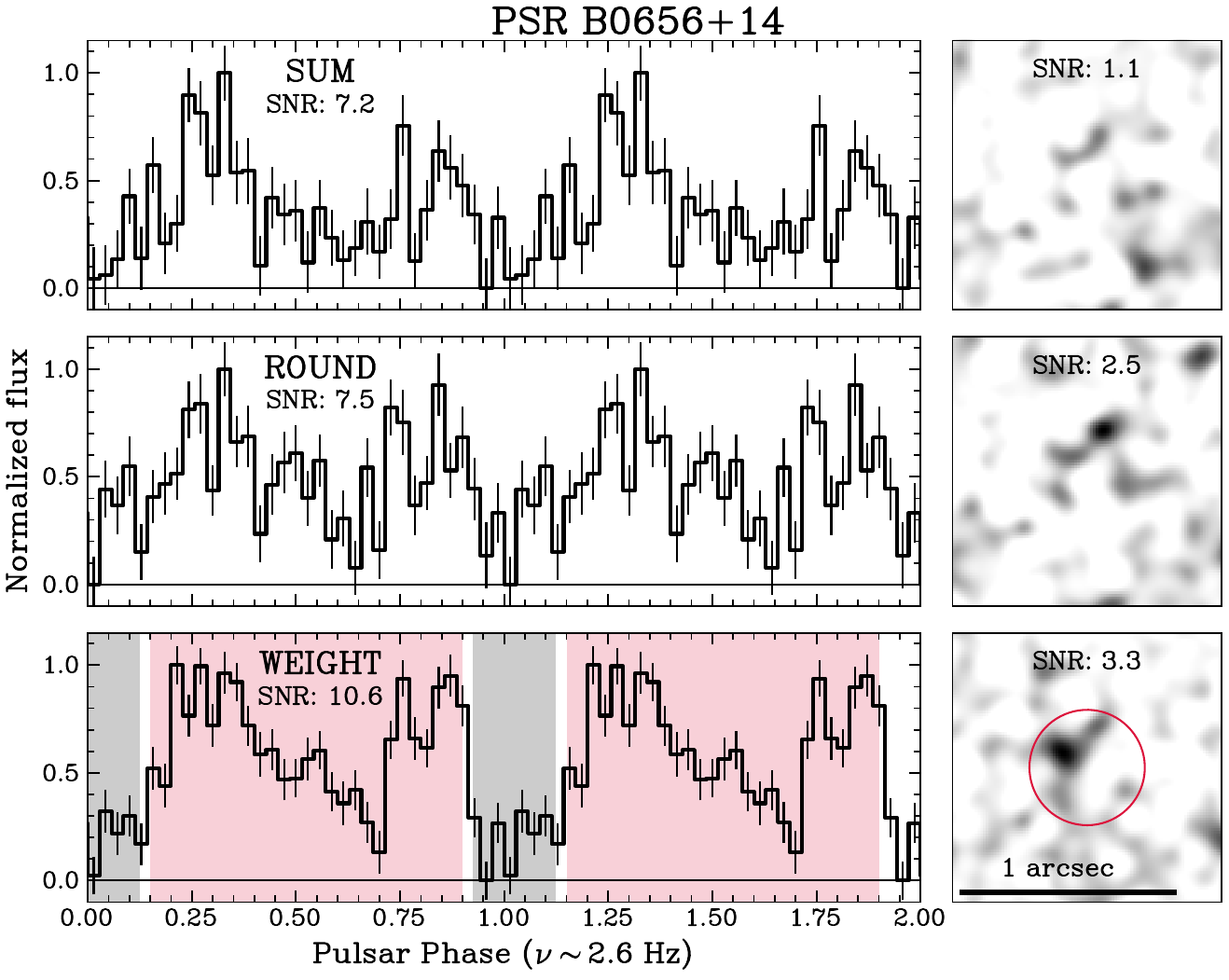}
    
    \caption{
        Normalized light curves (left) and on$-$off images (right) of B0540  ($V\sim 22.4$), Vela  ($V\sim 23.6$), and B0656 ($V\sim 25$) in descending order. The ``on'' (red) and ``off'' (gray) phases ranges are shown in the light curve. Each light curve is extracted with a different method, with the weight method introduced in this work being superior in all cases. 
    }
    \label{fig:pulsars}
\end{figure}

Phases are assigned to each collected frame using the \texttt{PINT} pulsar timing software \citep{luo2021pint}. The Crab ephemeris is obtained from the monthly Jodrell Bank observatory ephemerides, made accessible in the \texttt{CRABTIME} ephemeris catalog \citep{lyne1993years}. For B0540, we adopt the X-ray ephemeris collected by \citep{xie2024first} with the Imaging X-ray Polarimetry Explorer (IXPE). For Vela, we use the \textit{Fermi} Large Area Telescope (LAT) ephemeris from the third catalog of gamma-ray pulsars \citep{smith2023third}. For B0656, we use the UTMOST pulsar timing program  radio ephemeris \citep{jankowski2019utmost}. Using these phases, light curves are drawn from a $0''.6$-radius circular region surrounding the pulsar, whose position we measure from a long exposure image taken immediately prior. We also collect an ``on$-$off'' image, which is the difference between the summed frames collected in the pulsars' ``on'' and ``off'' phase ranges (chosen by hand). The on$-$off image is background subtracted, and we remove the influence of a handful of bad pixels with light sigma clipping.\footnote{Here we cut all pixels fluctuating by $>10\sigma$ from the image median, where $\sigma^2 = M_2 - M_1^2$, and $M_1/M_2$ are the median values of flux / flux squared in the image. This $\sigma$ formula was chosen to be comparable to the standard deviation (defined by replacing the medians with means), and using medians make the result less sensitive to outlier pixels.}

We also report SNRs for the light curve and on$-$off image. The light curve SNR is the amplitude divided by the mean flux uncertainty, and the image SNR is $F / (\sigma \sqrt{N})$, where $F$ is the flux inside the source region, $\sigma$ is the standard deviation in the background region, and $N$ is the number of pixels in the source region.  We smooth the image after computing the SNR for display purposes.

\subsection{Faint Pulsar Light Curves}
Fig.~\ref{fig:pulsars} shows the light curves and on$-$off images extracted for B0540, Vela, and B0656. The phase-averaged magnitudes quoted in the caption are obtained by \cite{zharikov2021psr} and references therein. The red / gray bands show the on / off phase ranges used to make the on$-$off images. The red circle in the on$-$off images represents the source region used to compute the SNR, and the area outside it is the background region. As proto-Lightspeed's image stability issues have now been mitigated \citep{layden_2026_spie}, new observations may deliver improved results by using weighted PSF photometry. Versions extracted using the ``sum,'' ``round,'' and ``weight'' methods are shown. Generally, our results are similar to previously published light curves \citep[see e.g.][]{cassanelli2025iqueye, kern2003optical}. 

B0540's weighted light curve is superior to the summed and rounded light curves as demonstrated by the $56\%$ increase in SNR compared to ``sum.'' Achieving the same SNR improvement purely by collecting more data would require $2.4\times$ as much observing time (assuming the SNR is proportional to the square root of observing time). Visually, the weighted light curve shape appears more realistic and is in agreement with the state-of-the-art light curve \citep{gradardi2011optical}. The on-off image also shows suppressed background noise, leading to a 57\% increase in image SNR. An earlier version of this light curve was previously published by \cite{layden2026proto}.

The success of the weighted method is even clearer in the Vela light curve. The sum and round method are not able to detect any of the pulses, whereas the two brightest are clearly detected in the weighted version, with possible detections of the third and fourth \citep[for a higher SNR light curve see][]{spolon2019timing}. This demonstrates that the weighting method is increasingly valuable for fainter sources. The pulsar detection in the image is less clear due to the image stability problems discussed in the previous section. However, some flux excess is clear within the red circle of Fig.~\ref{fig:pulsars}, which is most significant when the weight method is used.

Our B0656 also shows a dramatic light curve improvement when using the weight method. The light curve is now good enough to resolve possible structure in the two peaks, which were more coarsely sampled in the previous light curve \citep{kern2003optical}. A 47\% light curve SNR improvement is not as large as the improvement seen in B0540, possibly because noise in the ``sum'' light curve is boosting the apparent signal and therefore the SNR. A potentially more reliable metric is the significance of the pulse, as estimated using the $\chi^2$ of a constant-flux null hypothesis. This significance to which this null hypothesis is rejected rises from 6.1$\sigma$ for ``sum'' to 14$\sigma$ significance for ``weight.'' This 130\% SNR gain would require $\sim 5.3\times$ as much exposure time to accomplish with data alone.

\subsection{Other studies}

The ``weight" method is also valuable for stochastically varying, non-periodic sources. Since these light curves are not folded, each light curve bin contains only one $u$-value per pixel. The histogram of $u$-values for each pixel, which the weight method fits to, therefore contains only one event. Even with few $u$-values per light curve bin, we find that the weight method continues to exceed the other methods in precision. To demonstrate this, we analyze a 10 ms exposure proto-Lightspeed observation of the flaring X-ray binary GX 339$-$4, collected on September 13, 2025, with the ``sum,'' ``round,'' and ``weight'' methods. Initial analysis of these data were previously published by \cite{layden2026proto} and will be further discussed by \cite{berger2026}. A nearby (12$''$.5-distant) star was used to perform differential aperture photometry with \texttt{lightspeedpy}, with $2-4''$-radius annular background regions. The weight method provides a 51\% improvement over the other methods. The most significant advantage of higher SNR afforded by weighting is that it enables shorter frame rates, which can further resolve the faint, rapid flaring.

\begin{figure*}
    \centering
    \includegraphics[width=\linewidth]{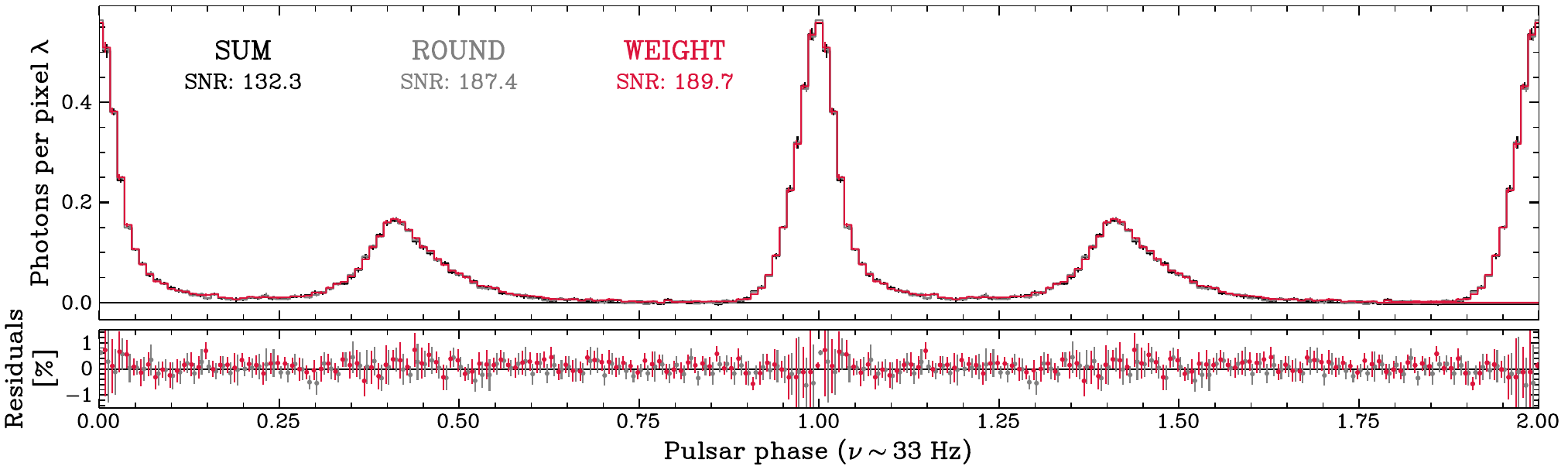}
    \caption{
       Light curves for the Crab pulsar extracted using the three methods discussed here. The vertical axis reflects the photon rate in counts per pixel frame at the center of the PSF. The bottom panel compares the round and weight method to sum. Detection is not difficult for this 16th magnitude source; rather, this figure shows that the weight method is unbiased as it produces identical results to the simpler methods when photons are plentiful. Also note that the weighted light curve is more significant, albeit less dramatically than for the fainter pulsars.
       \label{fig:crab}
    }
\end{figure*}

Finally, we show the Crab pulsar light curve in Fig.~\ref{fig:crab} to demonstrate that our use of the weight method is not producing biased results. At our 1.3 ms cadence, this $V=16.6$ pulsar delivers $\sim 0.6$ photons per frame for the brightest pixels at the center of the brighter peak, which is enough flux that all methods can easily detect it (though the weight method light curve has the largest SNR). Therefore, comparing the light curves serves to assess whether the methods provide consistently calibrated flux throughout the photon-starved regime. The residuals panel shows that the ``weight'' method matches the simpler ``sum'' method to within 1\% at all phases, which is within error bars. This confirms that the remaining modeling imperfections, such as the minor differences between the model $P(u|\lambda)$ distribution and the calibration data shown in Fig.~\ref{fig:likelihood}, do not noticeably bias the new weight method.

\section{Conclusions}
\label{sec:conclusion}
We have introduced a new ``weighted'' analysis method for sub-electron read noise detectors which enables more accurate analyses of faint sources. The method requires careful modeling of detector systematics such as the noise distribution and the flux-dependent internal QE. We perform for this modeling for proto-Lightspeed using calibration data and demonstrate the accuracy of the systematics models by showing that 1) we accurately reproduce the calibration data sets, and 2) fluxes estimated by the weight method are consistent with simpler methods when applied to a well measured source (the Crab pulsar).

We apply the weight method to light curves (folded and unfolded) and stacked images, and we find 50--130\% SNR increase compared to other methods with the faintest sources delivering the largest improvements. This method enabled significant proto-Lightspeed light curves for all four normal rotation-powered pulsars with confirmed optical pulses, and the improvement allowed us to use 2.3--5.3$\times$ less exposure time than other analysis methods would have required. Applied to aperiodic sources such as a flaring X-ray binary, the additional SNR could be used to shorten exposures and capture even more rapid astrophysical phenomena. Other astronomical observations of rapidly varying systems should also be improved by this method, such as transiting systems and faint high resolution mapping with lucky imaging.

We have made the analysis technique available through the \texttt{lightspeedpy} Python package. The weighted method does not pose much additional computational burden; for the analyses reported in this work, the weighted light curve extraction runs only $5-20\%$ slower than other methods, while for weighted photometry it runs $\sim 3\times$ slower. As an example of the absolute processing time, our 25 minutes of Vela data were processed in 10 minutes on a laptop.

Though we focus on an application to CMOS detectors and the proto-Lightspeed instrument, the principle of fitting a flux to the distribution of detected ADU counts applies to a wider array of detector technologies such as skipper CCDs and linear-mode avalanche photodiodes, allowing substantially improved photometry, astrometry, and imaging as more instruments enter the photon-counting regime.

\begin{acknowledgments}
This work was funded by the Kavli Foundation through the Kavli Institute Collaboration Kickstarter (KICK) grant PS-2025-GR-0251-3110. Proto-Lightspeed was funded by the MIT Kavli Research Investment Fund (grant MKI KRIF 2654352 and 2654373), by Kavli Institute Collaboration Kickstarter (KICK) Grant PS-2025-GR-0251-3110, and by the MIT John W. Jarve (1978) Seed Fund.
\end{acknowledgments}

\software{\texttt{PINT} \citep{luo2021pint}, \texttt{lightspeedpy} \citep{dinsmore2026lightspeedpy}, \texttt{Python}, \texttt{matplotlib}, \texttt{numpy}}
\facilities{Las Campanas Observatory}

\bibliography{bib}{}
\bibliographystyle{aasjournal}

\end{document}